\input epsf
\documentstyle[11pt]{article}
\def\a {\alpha}

\def\c {\gamma}
\def\d {\delta}
\def\l {\lambda}

\def\s {\sigma}
\def\t {\theta}
\def\w {\omega}
\def\C {\Gamma}
\def\D {\Delta}
\def\M {\cal M}
\def\T {\mbox{\boldmath$\tau$}}
\def\sig {\mbox{\boldmath$\sigma$}}
\def\L {\mbox{\boldmath$\Lambda$}}
\def\sel {{\cal X}_p}
\def\alg {{\cal F}_p}
\def\half {\frac{1}{2}}
\def\quarter {\frac{1}{4}}
\def\be{\begin{equation}}
\def\ee{\end{equation}}
\def\ba{\begin{eqnarray}}
\def\ea{\end{eqnarray}}

\def\R{{\mathchoice
{\setbox0=\hbox{$\displaystyle\rm R$}\hbox{\hbox to0pt
{\kern0.4\wd0\vrule height0.9\ht0\hss}\box0}}
{\setbox0=\hbox{$\textstyle\rm R$}\hbox{\hbox to0pt
{\kern0.4\wd0\vrule height0.9\ht0\hss}\box0}}
{\setbox0=\hbox{$\scriptstyle\rm R$}\hbox{\hbox to0pt
{\kern0.4\wd0\vrule height0.9\ht0\hss}\box0}}
{\setbox0=\hbox{$\scriptscriptstyle\rm R$}\hbox{\hbox to0pt
{\kern0.4\wd0\vrule height0.9\ht0\hss}\box0}}}}

\def\co{{\mathchoice
{\setbox0=\hbox{$\displaystyle\rm C$}\hbox{\hbox to0pt
{\kern0.4\wd0\vrule height0.5\ht0\hss}\box0}}
{\setbox0=\hbox{$\textstyle\rm C$}\hbox{\hbox to0pt
{\kern0.4\wd0\vrule height0.5\ht0\hss}\box0}}
{\setbox0=\hbox{$\scriptstyle\rm C$}\hbox{\hbox to0pt
{\kern0.4\wd0\vrule height0.5\ht0\hss}\box0}}
{\setbox0=\hbox{$\scriptscriptstyle\rm C$}\hbox{\hbox to0pt
{\kern0.4\wd0\vrule height0.5\ht0\hss}\box0}}}}

\def\Z{{\mathchoice
{\setbox0=\hbox{$\displaystyle\rm Z$}\hbox{\hbox to0pt
{\kern0.4\wd0\vrule height0.5\ht0\hss}\box0}}
{\setbox0=\hbox{$\textstyle\rm Z$}\hbox{\hbox to0pt
{\kern0.4\wd0\vrule height0.5\ht0\hss}\box0}}
{\setbox0=\hbox{$\scriptstyle\rm Z$}\hbox{\hbox to0pt
{\kern0.4\wd0\vrule height0.5\ht0\hss}\box0}}
{\setbox0=\hbox{$\scriptscriptstyle\rm Z$}\hbox{\hbox to0pt
{\kern0.4\wd0\vrule height0.5\ht0\hss}\box0}}}}

\def\sec#1{ \vskip7mm   \stepcounter{section}    \noindent
	{\large \bf \Roman{section}.  #1}
	\vskip3mm \setcounter{equation}{0}}
\def\subsec#1{ \vskip6mm  \stepcounter{subsection} \noindent
	{\bf \Roman{section}.\arabic{subsection}.  #1}
	\vskip3mm }

\begin{document}

\rightline{CGPG-95/3-1}
\rightline{gr-qc/9503064}
\vskip1.1cm
\centerline{\large \bf Non-covariance of the generalized holonomies: Examples}
\vskip8mm
 \renewcommand{\thefootnote}{\fnsymbol{footnote}}
\centerline{Troy A. Schilling\footnote{E-mail: troy@phys.psu.edu}}
 \setcounter{footnote}{0} \renewcommand{\thefootnote}{\arabic{footnote}}
\centerline{\it Center for Gravitational Physics and Geometry}
\centerline{\it Department of Physics, The Pennsylvania State University}
\centerline{\it University Park, Pennsylvania 16802,  USA}
\centerline{(October, 1994)}
\vskip8mm

\centerline{\bf Abstract}
\vskip.2cm
{\small
\begin{quotation}
\noindent  A key aspect of a recent proposal for a {\em generalized loop
representation}
of quantum Yang-Mills theory and gravity is considered.  Such a representation
of the quantum theory has been expected to arise via consideration
of a particular algebra of observables -- given by the traces of the
holonomies of {\em generalized loops}.  We notice, however,  a
technical subtlety, which prevents us from reaching the conclusion that the
generalized holonomies are covariant with respect to small gauge
transformations.  Further analysis is given which shows that they
are {\em not} covariant
with respect to small gauge transformations;  their traces are {\em not}
observables of the gauge theory.  This result indicates what may be
a serious complication to the use of generalized loops in physics.
\end{quotation}
}

\sec{Introduction}
There have recently been a variety of attempts to formulate gauge theories
in terms of loops \cite{ym}.
One of the key technical developments which suggests such a
formulation of gauge theory is Giles' result \cite{giles} that, for
$SU(N)$ theories, the information contained
in the Wilson loops (i.e. the traces of all holonomies around closed curves)
is sufficient for the reconstruction of the connection up to local gauge
transformations.  That is, the Wilson loops contain all of the gauge-invariant
information about the connection.  Since the Wilson loops
separate points of the space of connections modulo gauge transformations,
there is a sense in which any gauge-invariant function  on the
relevant space of connections (and hence any configuration observable of the
gauge theory) may be expressed in terms of the Wilson loops.
The (over-)completeness of these  observables suggests that they
be taken as the basic configuration observables in the quantization scheme.

This idea, along with Ashtekar's connection-dynamic formulation \cite{ashtekar}
of general relativity, provide the foundation of the Ashtekar-Rovelli-Smolin
approach to quantum gravity.   The duality between connections and loops
\cite{ash-lew} suggests the possibility of representing states by
functions
of loops.  The idea of the loop representation was first introduced to
gravity by Rovelli and Smolin \cite{smolin}, and has resulted in a formalism
with several attractive features.  Most notable are the relationship between
diffeomorphism invariance and knot theory, the combinatorial aspect of the
formalism, and a sense in which discreteness emerges.
For details of this approach to quantum gravity see \cite{lectures}.

Despite the merits of the loop-representation, it is generally believed that
ordinary loops are not sufficient for the description of gravity.
The problem of regularization of the Wilson loop operators suggests that
one introduce a thickening, or framing, of the loops.
A novel suggestion is presented  by  recent results of
Di Bartolo, Gambini and Griego \cite{gambini, regularization}.
The space of loops based at a fixed point forms a group \cite{trias},
but not a Lie group.  In an attempt to ``coordinatize'' the group of based
loops,  the authors of \cite{gambini} came upon a
generalization of the notion of a loop.  The result is an infinite-dimensional
Lie group, which contains the group of (based) loops as a subgroup.
The elements of this {\em extended loop group}\footnote{
	Note that the ``extended loop group'' and the ``group of loops''
	 discussed below are unrelated to  what mathematicians call the
	``loop group''.  }
are (sequences of) distributional
quantities,  ordinary loops being the ``most distributional''
elements.
There are also elements, however, which are ``less distributional'' than
loops,  objects that we may think of as ``smoothened loops''.  With the
above regularization issue in mind,
the existence of the smoothened loops is of obvious
interest.  An anticipated benefit of the use of generalized loops is the
ability to apply familiar functional methods to the study of the (generalized)
loop representation.  Further, we will see that  the extended loop group
has the global structure of a vector space.  Hence, integration on the
generalized loop space is a fairly straight-forward matter.
Integration techniques may supply a form of the inner-product which
is inherent to the loop space.

For these reasons, one is motivated to examine the role of generalized
loops in gauge theory.  Recall that the idea of the loop representation is
based on the Wilson loops.  Since, as we will see,
the generalized loops are defined via
inspection of the functional form of the holonomies of ordinary loops,
the holonomy formally extends to the extended loop group.
It is through this extension of the holonomy that one can imagine the
construction of a generalized loop representation.
For such a formulation of quantum
Yang-Mills theory or gravity to make sense, the generalized
holonomies  must be
gauge-covariant with respect to (small) gauge transformations.  At the very
least, the traced holonomies should be gauge-invariant.  The main result
presented here is the fact that {\em the generalized holonomies are not
covariant with respect to small gauge transformations}.  Despite the beauty
of the extended loop group, its use in gauge theory may therefore be
limited.

In Sec. II, we review the construction of the extended loop
group and discuss some useful properties of its elements.  Section
III focuses on the generalized holonomies, with  particular attention given
to their transformation properties.  Consideration of the Abelian case
will suggest simple examples of non-covariance of the generalized $SU(2)$
holonomies.
Two such examples are presented in Sec. IV.  Finally, in Sec. V,
we conclude with generalizations of the results and remarks concerning their
relevance in physics.

\sec{Generalized loops}
The purpose of this section is to recall the basic ideas regarding
generalized loops (for details, see the original work \cite{gambini}).
After introducing
the group of loops on an arbitrary connected manifold, consideration of
the holonomies will suggest a generalization of the notion of loops.
The set of these generalized loops -- the  extended loop group -- forms an
infinite dimensional Lie group.

Fix a point $p$ on an arbitrary connected manifold $\M$ and consider the space
${\cal C}_{p}$ of closed curves based at $p$.  Elements of ${\cal C}_{p}$ are
piecewise-smooth
maps $C:I \rightarrow \M$ such that $C(0) = p = C(1)$, where $I$ is the unit
interval.  Now, our main motivation for such consideration is that the trace
of the holonomy of a physical gauge field around any closed curve is
gauge-invariant; i.e. an observable of the classical field theory.
Since we are primarily concerned with the observables,
we are not  interested in the space ${\cal C}_{p}$ itself,
but in a space of equivalence classes of elements of ${\cal C}_{p}$.
For example, two closed curves which differ merely by reparametrization
yield the same holonomies for an arbitrary smooth connection over $\M$.
Gambini and
Trias \cite{trias} provide an  appropriate identification
of elements of ${\cal C}_{p}$:  Two closed curves $C, C^\prime
\! \in \! {\cal C}_{p}$ are deemed equivalent if
$C \circ \bar{C^\prime}$ is
contractible {\em within itself} to the trivial curve $\iota(s) \equiv p$,
where $\bar{C^\prime}$ is the reversed path $\bar{C^\prime}(s) =
{C^\prime}(1-s)$.
With this equivalence relation on
${\cal C}_{p}$, it is easy to see that two closed curves which are equivalent
give the same holonomies for any smooth connection over $\M$.  Denote by
${\cal L}_{p}$ the space obtained by dividing ${\cal C}_{p}$ by this
equivalence relation.  The obvious composition
of paths induces a {\em group operation} on ${\cal L}_{p}$.

Next, consider a connection ${\bf A}$ on a principle bundle $P({\M}, G)$,
where $G$ is a compact, connected Lie group.  For the sake of simplicity,
we shall assume that $P$ is trivial and view connections as Lie algebra-valued
one-forms on $\M$.  (Below, we will restrict attention to the case $G=SU(2)$,
for which every bundle is trivial.)
To an element
$\c \! \in \! {\cal L}_{p}$ we may associate the holonomy, $U_{A}[\c]$, around
any path $C$ in the equivalence class defining $\c$.  Expressed in terms
of the fundamental representations of the gauge group $G$ and its Lie
algebra, the holonomy takes the form of the path-ordered exponential,
which may be written explicitly as
\ba
U_{A}[\c] &=& P \exp\oint\limits_C {\bf A}		\nonumber	\\
	  &=& \sum_{n=0}^{\infty}{\int} {\cdots} {\int}
	\, X_{\c}^{a_1{\cdots}a_n}(x_1, {\ldots}, x_n) \,
	{\bf A}_{a_1}(x_1){\cdots}{\bf A}_{a_n}(x_n),		\label{holon}
\ea
where
\be
X_{\c}^{a_1{\cdots}a_n}(x_1, {\ldots}, x_n) :=
   \oint\limits_{C}dy^{a_1}\int_{y_{1}}^1dy^{a_2}{\cdots}\int_{y_{n-1}}^1
	dy^{a_n} \,  {\d}(x_1, \, y_1){\cdots}{\d}(x_n, \, y_n),
							\label{Xdefn}
\ee
and the zeroth term in Eq. (\ref{holon}) is taken to be the identity.
In Eq.~(\ref{Xdefn}) the index $a_{k}$ is ``attached'' to the point
$x_k$, and for each $n$, $X_{\c}^{a_1{\cdots}a_n}(x_1, {\ldots}, x_n)$
is an $n$-point distributional vector density of weight one in
each argument $x_k$.
As suggested by the subscript on the $X$'s, these $n$-point
distributions are independent of the particular path $C$ chosen from the
equivalence class determined by $\c \! \in \! {\cal L}_{p}$.
It will be convenient to employ the notation
\be
	X_{\c}^{{\mu}_{1} {\cdots} {\mu}_{n}} :=
	X_{\c}^{a_1{\cdots}a_n}(x_1, {\ldots}, x_n),	\label{notation}
\ee
where the index ${\mu}_{k}$ now represents the pair $(a_{k}, x_{k})$, and
contraction of greek indices represents both contraction of the latin
index and the integration over $\M$.

Thus, to every element $\c \! \in \! {\cal L}_{p}$ is associated a string
\[
X_{\c} := (1, \, X_{\c}^{{\mu}_1},{\ldots}, \, X_{\c}^{{\mu}_1{\cdots}{\mu}_n},
					{\ldots})
\]
of multi-vector densities.  As is observed in  \cite{gambini}, if
$\c, \eta \! \in \! {\cal L}_{p}$, the multi-densities corresponding to their
product may be expressed as
\be
X_{\c \circ \eta}^{{\mu}_1 {\cdots} {\mu}_n} =
	\sum_{k=0}^nX_{\c}^{{\mu}_1 {\cdots} {\mu}_k}
		 X_{\eta}^{{\mu}_{k+1} {\cdots} {\mu}_n},	\label{prod}
\ee
with the convention that
\[
	X^{{\mu}_1 {\cdots} {\mu}_0} := 1.
\]

These strings of multi-densities satisfy two useful identities.  Denote
by $X$ the string corresponding to an arbitrary loop in ${\cal L}_p$.
The first identity reflects the ordering of points on the image of the loop;
\be
X^{\underline{{\mu}_1 {\cdots} {\mu}_k} {\mu}_{k+1} {\cdots} {\mu}_n}
	= X^{{\mu}_1 {\cdots} {\mu}_k} X^{{\mu}_{k+1} {\cdots} {\mu}_n},
								\label{alg}
\ee
where the left side is obtained by summing over all permutations
of the ${\mu}_i$ which preserve the relative ordering of the first $k$
indices and also the relative ordering of remaining $n\!-\!k$.  For example,
\[
 X^{\underline{{\mu}_1 {\mu}_2} {\mu}_3 {\mu}_4} =
X^{{\mu}_1 {\mu}_2 {\mu}_3 {\mu}_4} + X^{{\mu}_1 {\mu}_3 {\mu}_2 {\mu}_4} +
X^{{\mu}_1 {\mu}_3 {\mu}_4 {\mu}_2} + X^{{\mu}_3 {\mu}_1 {\mu}_2 {\mu}_4} +
X^{{\mu}_3 {\mu}_1 {\mu}_4 {\mu}_2} + X^{{\mu}_3 {\mu}_4 {\mu}_1 {\mu}_2}.
\]
Next, since taking the divergence of a vector density requires no additional
structure (e.g. a metric or derivative operator) on $\M$, it is
natural to ask whether the
divergence of an entry of $X$ satisfies any useful property.  The answer
is in the affirmative;
\be
 \frac{\partial}{{\partial}x^{a_k}} X^{{\mu}_1 {\cdots} {\mu}_n} =
 [\d(x_k, x_{k-1}) - \d(x_k, x_{k+1})]  \,
	X^{{\mu}_1 {\cdots} {\hat{\mu}}_k {\cdots} {\mu}_n},	\label{diff}
\ee
where the caret over the ${\mu}_k$ is intended to indicate its absence,
and the mixed notation on the left side should be transparent.
By definition, $x_0$ and $x_{n+1}$ are taken to be the base point, $p$.
Thus the divergence of the rank-n entry of $X$ is directly related to the
rank-(n-1) entry.

The basic idea of Di Bartolo, Gambini and Griego \cite{gambini} is to
consider the space
of {\em all} objects satisfying these relations.  To be precise, let
${\cal E}$ be the space of all sequences \linebreak
$(E^0, E^{\mu_1}, {\ldots},
E^{{\mu}_1 {\cdots} {\mu}_n}, {\ldots})$, where $E^0$ is a real number
and $E^{{\mu}_1 {\cdots} {\mu}_n}$ is a {\em distributional} vector-density
in each index.  ${\cal E}$ becomes an associative algebra when equipped with
the product motivated by Eq.~(\ref{prod}); given $E_1, E_2 \! \in \!
{\cal E}$, we define
\be
 (E_1 \times E_2)^{{\mu}_1 {\cdots} {\mu}_n} :=
   	\sum_{k=0}^nE_1^{{\mu}_1 {\cdots} {\mu}_k}		\label{assoc}
		 E_2^{{\mu}_{k+1} {\cdots} {\mu}_n},
\ee
where $E^{{\mu}_1 {\cdots} {\mu}_0} := E^0 \! \in \! \R$.
The extended loop group (based at $p \! \in \! \M$) is defined to consist of
those elements of ${\cal E}$ which satisfy the algebraic relation~(\ref{alg})
and the differential relation~(\ref{diff}) and for which the rank-zero entry
is unity.  This set, denoted by $\sel$, is closed under the product defined
above and every element is seen to have an inverse with respect to the
identity element, ${\rm I}:=(1, 0, 0 {\ldots})$.  $\sel$ is then a group which
contains ${\cal L}_p$ as a subgroup.

Next, $\sel$ is an infinite-dimensional Lie group in
the following sense.  For any element $X \! \in \! \sel$, the logarithm
\[
 \ln(X) := \sum_{m=1}^\infty \frac{(-)^{m+1}}{m} (X-{\rm I})^m
\]
is a well-defined element of ${\cal E}$ (with vanishing rank-zero entry)
which satisfies the differential
relation given by Eq.~(\ref{diff}) and the {\em homogeneous
algebraic relation}
\be
 \ln(X)^{\underline{{\mu}_1 {\cdots} {\mu}_k} {\mu}_{k+1} {\cdots} {\mu}_n}
  = 0 \quad \forall \quad 0<k<n.				\label{hom-alg}
\ee
Let $\alg$ consist of all elements of ${\cal E}$ which satisfy these two
conditions and with vanishing rank-zero entry.  One can show that
 if $F \! \in \! \alg$ then $\exp(F) := \sum_{k=0}^{\infty} \frac{1}{k!}F^k$ is
   a well-defined element of $\sel$.
Further, for any element $X \! \in \! \sel$ the logarithm $F = \ln X$ is the
unique element of $\alg$ for which
   $X = \exp(F)$.
$\alg$ is closed under the Lie bracket given by the commutator with
respect to the associative product (\ref{assoc}), defined on $\cal E$;
this is the Lie bracket relevant to the group operation on $\sel$.
Thus, $\alg$ is simply the Lie algebra corresponding to $\sel$.
Note also that $\sel$ has the global structure of an infinite-dimensional
vector space since there is a one-one correspondence between its elements and
elements of its Lie algebra.  In particular, we may unambiguously take the
real power of any element in $\sel$;  $X^t := \exp (t\ln X)$.

Given a  $G$-connection, one may consider the formal expression for the
holonomy around an arbitrary extended loop,
\be
 U_A[X] := \sum_{n=0}^{\infty} X^{{\mu_1} {\cdots} {\mu}_n} \,
    {\bf A}_{{\mu}_1} {\cdots} {\bf A}_{{\mu}_n} .  	\label{holonomy}
\ee
There is no claim that the extended holonomies take
values in the gauge group, or even that they converge.  However, in
\cite{gambini} it is formally shown that
$U_A[X_1 \times X_2] = U_A[X_1] U_A[X_2]$, where the right side is given
by matrix multiplication in the fundamental representation.
At least at the formal level, the holonomy
extends to a homomorphism on $\sel$.
It is worth noticing one particular situation
in which the extended holonomies converge to elements of the gauge group.
Suppose that ${\bf A}$ is an {\em Abelian} connection; i.e.
$[{\bf A}(x), {\bf A}(y)] \equiv 0$.
Then, using the algebraic relation (\ref{alg}),
it is a simple matter to show that for any $X \! \in \! \sel$
\ba
 U_A^{(n)}[X] &:=& X^{{\mu}_1 {\cdots} {\mu}_n} {\bf A}_{{\mu}_1}
	{\cdots} {\bf A}_{{\mu}_n}
								\nonumber   \\
	&=& \frac{1}{n!} (X^{\mu}{\bf A}_{\mu})^n.		\label{abelian}
\ea
So the extended holonomy corresponding to any Abelian connection is just
given by the exponential of $X^{\mu} {\bf A}_{\mu}$.
(The result (\ref{abelian}) depends only on
the fact that the restriction of ${\bf A}$ to the support of $X$ is Abelian.)
If, for example, the
support of ${\bf A}$ is also of compact closure, the holonomy is convergent
and group-valued on all of $\sel$.
This result will be used extensively in what
follows.

\sec{The generalized holonomies}

The construction of the extended loop group is elegant and of considerable
interest from a purely mathematical point of view.
However, the intention extends to physics as well.
The idea is simply to generalize the formalism used in the ordinary loop
representations of gauge theories, i.e. to consider the traces
of the extended holonomies as observables for Yang-Mills theory
and, perhaps more importantly, general relativity.  In particular,
an extended loop
representation for
quantum general relativity may be an especially useful setting for
consideration of an inverse loop transform and the framing
problem of knot invariants \cite{regularization, reg2}.

With the intended application of the generalized loops in mind, it is
natural to examine the behavior of the extended holonomies under gauge
transformations.  One often
distinguishes between two types of gauge freedom.  The gauge
which is generated by the (first-class) constraints is {\em physical} gauge
freedom, while that which is not is {\em symmetry}.   The physical gauge
freedom then corresponds to that generated by the infinitesimal gauge
transformations.  Thus, in order for the traces of the extended holonomies
to  give observables, it is necessary that they be
invariant under small gauge transformations.
This issue was considered in \cite{gambini}, but as is usual in pioneering
work, a detailed analysis was sacrificed for the sake of progress in other
directions.

Such an analysis is the purpose of this section.
We will find that there is a technical subtlety which prevents
us from accepting the naive conclusion that the extended holonomies
are (formally) gauge covariant with respect to infinitesimal gauge
transformations.  In order to gain some insight about the transformations
properties of the holonomies, we will then consider the Abelian case.
In all that follows, we will restrict attention to the manifold
${\M} \approx {\R}^{3}$.

Recall that an infinitesimal gauge transformation
is given by a map $\L : {\M} \rightarrow {\cal L}G$, where ${\cal L}G$
is the Lie algebra of $G$.  To first order in $\L$, the
gauge-transformed connection is given by
\[
 {\bf A}^{\Lambda} = {\bf A} + d\L + [{\bf A}, \L].
\]
Set
\[
 U_A^{(n)}[X] := X^{{\mu}_1 {\cdots} {\mu}_n} {\bf A}_{{\mu}_1} {\cdots} {\bf
A}_{{\mu}_n}
\]
as in Eq. (\ref{abelian}), so that
\[
 U_A[X] = \sum_{n=0}^{\infty} U_A^{(n)}[X].
\]
Using the differential relation (\ref{diff}) satisfied by the generalized
loops,
one may obtain the holonomy corresponding to the gauge-transformed connection;
\be
 U_{A^{\Lambda}}^{(n)}[X] = U_A^{(n)}[X] + \left[ U_A^{(n-1)}[X], \,
	{\L}(p) \right]
		+ f_{(A,\Lambda)}^{(n)}[X] - f_{(A,\Lambda)}^{(n-1)}[X] ,
								\label{transf}
\ee
where
\be
 f_{(A,\Lambda)}^{(n)}[X] := \sum_{k=1}^n  X^{{\mu}_1 {\cdots} {\mu}_n}
	{\bf A}_{{\mu}_1} {\cdots} {\bf A}_{{\mu}_{k-1}}[{\bf A}, \L]_{{\mu}_k}
	{\bf A}_{{\mu}_{k+1}} {\cdots} {\bf A}_{{\mu}_n}.	\label{f}
\ee

One is tempted to conclude from Eq.~(\ref{transf}) that, since the
$f^{(n)}$ cancel upon summation of the series for the transformed holonomy,
all extended holonomies are formally gauge-covariant.
However, let us proceed more carefully.  Consider the partial sum,
\be
 \sum_{n=0}^N U_{A^{\Lambda}}^{(n)}[X]=\sum_{n=0}^{N} U_A^{(n)}[X] +
  \left[ \sum_{n=0}^{N} U_A^{(n)}[X], \L(p) \right] - \left[U_A^{(N)}[X],
  \L(p) \right]   +   f_{(A,\Lambda)}^{(N)}[X].			\label{part-sum}
\ee
If we suppose that $U_A[X]$ converges, then $U_A^{(N)}[X] \! \rightarrow \! 0$
as $N \! \rightarrow \! \infty$ and  the extended holonomy is covariant
under infinitesimal  gauge-transformations only if
\be
 f_{(A,\Lambda)}^{(N)}[X]\rightarrow 0 \quad\mbox{as}\quad N \rightarrow\infty.
\ee
Thus,  gauge-covariance of the extended holonomies does not follow trivially
from Eq.~(\ref{transf}).

We are now faced with the problem of whether the notion of holonomy generalizes
to the extended loop group.  Although, by inspection of Eq. (\ref{f}), the
Abelian case is trivial, an example will lead the way to an
understanding of the non-Abelian case.  Therefore, let us consider $G = U(1)$.
Let $\c$ be the loop determined by the curve
\[
 C(s) = \left( \cos(2\pi s), \, \sin(2\pi s), \, 0 \right).
\]
This loop determines an element $X_{\c} \! \in \! \sel$, where the base point
has been fixed as $p = (1, 0, 0)$.  We will focus on the generalized holonomies
of an arbitrary real power, $X_{\c}^t$, of this particular loop.

The generic $U(1)$-connection is of the form
\[
 {\bf A}(x) = -i{\omega}(x),
\]
where ${\w}$ is a real one-form on $\M$.  To compute the holonomy of
${\bf A}$ around $X_{\c}^t$, we need only know the rank-1 entry of
$X^t_{\c}$,
\[
 (X_{\c}^t)^\mu  =  t \, X_{\c}^\mu,
\]
and that, with respect to the cylindrical coordinates $z, r, \t$,
\[
 X_{\c}^\mu  =  \frac{1}{r} \, {\d}^1(r, 1) \, {\d}^1(z, 0)
	\left( \frac{\partial}{\partial \t} \right)^\mu \,\, .
\]
By Eq.~(\ref{abelian}), the holonomy is then given by
\be
 U_A[X_{\c}^t] = \exp(t X^{\mu} {\bf A}_{\mu}) = \exp( -it \oint_{\C} {\w} ),
							     \label{u1-holon}
\ee
where $\C$ is the unit circle in the $x$-$y$ plane (the image of $C$).
The holonomy of the gauge-transformed connection
${\bf A}^g = {\bf A} + g^{-1} {\bf d}g$ is
\be
  U_{A^g}[X_{\c}^t] = U_A[X_{\c}^t] \cdot \exp( -2{\pi}it \, w_{\c}[g]),
								\label{u1}
\ee
where
\be
  w_{\c}[g] := \frac{i}{2\pi} \oint_{\C} g^{-1} \, dg \! \in \! \Z.
\ee

Note that the pull-back of $g$ to $\Gamma$ is a map from the circle
into $U(1)$ and that $w_{\c}[g]$ is simply the winding number of this map, i.e.
the number of times $g$ wraps $U(1)$ around  $\C$.
Suppose $g$ is a small
gauge transformation.  Then, by definition,
there exists a homotopy $g_{\l}$
(a smooth one-parameter family of gauge transformations, $\l \! \in \! [0,1]$)
connecting $g = g_1$ to the trivial map $g_0 \equiv {\bf 1}$.
By pulling the homotopy back to $\C$,
one then obtains a one-parameter family of maps
from the circle into $U(1)$.  Since the winding number is integral, it
must be the same for each $g_{\l}$.  Hence,
$w_{\c}[g] = w_{\c}[g_1] = w_{\c}[g_0] = 0$.  We then see that the
holonomy (\ref{u1-holon}) is covariant with respect to small gauge
transformations.  Note that this was not a general proof of covariance of the
$U(1)$-holonomies (the general proof if much simpler that what we have done
above!).  The above reasoning applies only to the arbitrary real power of
the particular loop $\c$.  The utility of our result, however,  lies not in the
conclusion, but in the methodology.  The above ansatz, when applied to the case
$G = SU(2)$, will suggest simple examples which show that the non-Abelian
holonomies are {\em not} covariant with respect to small gauge.

\sec{Non-covariance of the generalized holonomies}
We can study the non-Abelian case, by ``embedding'' the above result into
$SU(2)$.  The idea in mind is to replace $U(1)$  by an Abelian subgroup
of $SU(2)$.  After making this idea more precise, natural examples of
non-covariance will be presented.  The first is, perhaps, the most natural;
it involves the holonomy of the real power of an ordinary loop.  For the
second example, we will consider the holonomies of generalized loops which are
``least distributional'', in a sense to be explained below.

Let us consider an arbitrary Abelian subgroup of $SU(2)$.  This subgroup is
generated by an element, ${\bf T}$, of the Lie algebra, ${\cal L}SU(2)$.
We may assume, without loss of generality, that ${\bf T}$ is normalized as
\[
	\mbox{tr}({\bf T}^2) = -2,
\]
so that, for example,
\be
  \exp[r{\bf T}] = \exp[(r + 2 \pi n){\bf T}] \quad \forall \quad
	n \! \in \! \Z,	\: r \! \in \! \R.		\label{norm}
\ee
Now suppose ${\bf A}$ is an $SU(2)$-connection which
is proportional to ${\bf T}$; i.e.
\be
	{\bf A} = {\w} {\bf T}
								\label{conn}
\ee
for some one-form ${\w}$.
Suppose further that $g : {\R}^3 \rightarrow SU(2)$ is an $SU(2)$ gauge
transformation whose restriction to $\C$ is contained in the $U(1)$-subgroup
generated by ${\bf T}$.  We can now mimic the discussion leading to
Eq.~(\ref{u1}) by making the replacement $-i \mapsto {\bf T}$.
We obtain
\be
  U_{A^g}[X_{\c}^t] = U_A[X_{\c}^t] \cdot \exp(2 \pi t v_{\c}[g] {\bf T}),
								\label{su2}
\ee
where $v_{\c}$ is defined as
\be
  v_{\c}[g] {\bf T} := \frac{1}{2 \pi} \, \oint_{\C} g^{-1}{\bf d}g.
								\label{v}
\ee
The meaning of $v_{\c}$ is analogous to that of $w_{\c}$; it is simply the
number of times $g$ winds the $U(1)$-subgroup generated by ${\bf T}$
around the circle $\C$.  Of course, $v_{\c}$ is only defined for such an
Abelian gauge transformation.

For integral $t$, $X_{\c}^t$ corresponds to an ordinary loop and the holonomy
is covariant under all gauge transformations.  But for the above holonomy to
transform covariantly for all {\em real} $t$, $v_{\c}[g]$ must be trivial.
Recall that in Sec. III it was the non-simple connectivity of $U(1)$ that
prevented us from finding small gauge transformations with non-trivial winding
number around $\C$.  $SU(2)$ is, of course, simply connected; hence, there
is no immediate suspicion that there do not exist small gauge transformations
with nontrivial $v_{\c}$.  In fact, there do exist such gauge transformations.
The task at hand is to produce an explicit expression for a small gauge
transformation whose restriction to $\C$ lies is an Abelian subgroup of
$SU(2)$, and which winds $\C$ non-trivially around this Abelian subgroup.
Since the exponential factor in Eq. (\ref{v}) is independent of the connection,
we will then have shown that the generalized holonomies are not gauge
covariant.

To this end, let us first focus on a convenient description of the
manifold structure of $SU(2)$.
The Lie algebra, ${\cal L}SU(2)$, of $SU(2)$ is
a real, three-dimensional vector space with a natural (Killing-Cartan)
inner-product
which, in the fundamental representation, takes the form\footnote{
	The usual factor of $i$ has been absorbed into the definition
	of the Lie algebra elements.  With this convention, ${\cal L}SU(2)$
	is represented by traceless {\em anti}-Hermitian matrices.
	This eliminates annoying powers of
	$i$ which otherwise would have appeared in Eq.~(\ref{holon}).}
\be
	({\bf T}_1, {\bf T}_2) := -2 \, \mbox{tr}({\bf T}_1{\bf T}_2).
\ee
Fix a basis $\{ {\T}_1, {\T}_2, {\T}_3 \}$ for the Lie algebra, which is
ortho-normal with respect to this inner-product.  For any element
${\bf L} \! \in \! {\cal L}SU(2)$, $\exp{\bf L}$ may be uniquely written
as\footnote{
	The common example is obtained by choosing ${\T}_i
	=(-i/2){\sig}_i$, where ${\sig}_i$ are the
	Pauli matrices.}
\be
	  \exp({\bf L}) = a_0{\bf 1} + 2[a_1{\T}_1 + a_2{\T}_2 + a_3{\T}_3],
								\label{sphere}
\ee
where $a_0^2 + a_1^2 + a_2^2 + a_3^2 = 1$.  While  $a_1, \ldots ,a_4$ may be
written in terms of ${\bf L}$, we will not find their explicit form useful.
Having chosen a basis $\{ {\T}_1,{\T}_2, {\T}_3 \}$ for the Lie
algebra, we then obtain an isomorphism of $SU(2)$ with the unit 3-sphere
in ${\R}^4$.  Of course, the identity {\bf 1} is represented by the point
$(1, 0, 0, 0)$.  We will abuse notation  and write
$\exp({\bf L}) = (a_0, a_1, a_2, a_3) = a_i \vec{e}_i$,
where the $a_i$ are those appearing in Eq. (\ref{sphere}), and
$\vec{e}_i$ form the obvious ortho-normal basis of ${\R}^4$.
The algebra ${\cal L}SU(2)$ may be viewed as the tangent space to
$SU(2)$ at the identity, and with the Lie algebra element, ${\bf L} =
L^1{\T}_1 + L^2{\T}_2 + L^3{\T}_3$, we may identify the vector
$(0, L^1, L^2, L^3) \in {\R}^4$.
The $U(1)$-subgroup generated by ${\bf L}$ is now simply represented by the
great circle (through ${\bf 1}$) whose tangent at the identity is
proportional to ${\bf L}$.
For example, $\exp(\a {\T}_1) = ( \cos(\a/2), \sin(\a/2), 0, 0)$.

We may now view a gauge transformation as a smooth
map $g: {\R}^3 \rightarrow {\R}^4$ whose image is contained in the unit
3-sphere.
Choose the basis $\{ {\T}_1, {\T}_2, {\T}_3 \}$ so that the (arbitrary)
algebra element considered above is given by
${\bf T} = 2{\T}_1$  and let us look for a small gauge transformation
$g$ for which $g(\cos \t, \sin \t, 0) = \exp({\bf \t T}) =
(\cos \t, \sin \t, 0, 0)$.  We will then have $v_{\c}[g] = 1$, and our
goal will have been accomplished.

At this point, a brief digression will be quite instructive.
Let us  display a particular homotopy
connecting the curve $h(\t) = \exp( \t {\bf T})$ to the trivial
curve $\iota(\t) \equiv {\bf 1}$.  This can be done geometrically, as follows.
Consider the intersection of $SU(2)$ (the 3-sphere in ${\R}^4$) and
the hyperplane $P^3 = \{ \vec{a} \! \in \! {\R}^4 | a_3 = 0\}$.  This is a
2-sphere in ${\R}^4$, which we will denote as $S$.  Let $P^2(\a)$
be the 2-plane in $P^3$ consisting of points of the form
$\vec{e}_0 + \vec{v}$ such that $\vec{v} \cdot \vec{n}(\a) = 0$,
where $\vec{n}(\a) = \vec{e}_2 \cos \a + \vec{e}_0 \sin \a$.
The intersection of $S$ with $P^2(\a)$ is a circle of radius
$r(\a) = \sin \a$, which may be parameterized as
\be
  h_{\a}(\t) = \left(
  \begin{array}{c}
	1 - (1 - \cos\t)\sin^2\a 	\vspace{1mm} \\
	\sin\t \sin\a			\vspace{1mm} \\
	(1 - \cos\t)\sin\a \cos\a 	\vspace{1mm} \\
	0
  \end{array}
       		\right),
\ee
for $\t  \in [0, 2\pi]$.
As $\a$ varies from $0$ to $\pi$, these circles ``foliate'' the sphere $S$.
Notice, in particular,  that
$h_0(\t) \equiv {\bf 1} \,\, \mbox{and} \,\, h_{\pi /2}(\t) =
(\cos \t, \sin \t, 0, 0) = \exp (\t {\bf T})$.
Therefore, $h_{\a}$ provides the desired homotopy (see Fig. 1.)
This homotopy will play a very important role in the examples that follow.

\begin{figure}[t]
\vskip-8mm
\epsfysize=2.3in
\hspace{5.7cm}   \epsffile{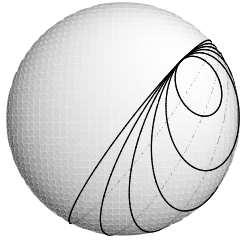}
\vskip-8mm
\caption{\protect\footnotesize A convenient homotopy connecting the trivial
	curve to the curve winding once around
	the $U(1)$-subgroup generated by ${\bf T}$.  For each $0 \le \a \le
	\pi/2$,
	the image of $h_{\a}$ is a circle on the 2-sphere $S$, as shown
	for various values of the parameter ${\a}$.
	As $\a$
	decreases from $\pi/2$ to $0$, $h_{\a}$ shrinks to a point
	(the identity element).}
\end{figure}


\subsec{Example 1}
We can now suggest the form of a gauge
transformation $g : {\R}^3 \rightarrow SU(2)$ which is demonstrably small,
whose restriction to the unit
circle in the $x$-$y$ plane lies in the Abelian subgroup generated by {\bf T},
and which winds this subgroup non-trivially around the unit circle.
In order to satisfy generic boundary conditions at infinity, we will
also demand that $g$ be trivial outside of a compact region.

The desired gauge transformation may be obtained from a smooth assignment
of the angle $\a$ to each pair of cylindrical coordinates $r, z$;
i.e. we try
\be
	g(r\cos \t, r\sin \t, z) := h_{\a (r,z)}(\t),
\ee
such that $\a (1, 0) = \pi/2$.
Any such gauge transformation is obviously small since the one-parameter
family of gauge transformations
\be
  g_{\l}(r\cos \t, r\sin \t, z) := h_{\l \a (r,z)}(\t), \quad \l \in [0,1]
\ee
provides a homotopy connecting $g$ to the trivial map.

It remains only to produce the assignment $\a (r, z)$.  This may be
accomplished by use of the smearing function
\be
    {\s}_{\Delta}(x) := \left\{
	\begin{array}{c @ {\quad : \quad} l}
	 \mbox{e} \cdot\exp\!\left( \frac{-{\Delta}^2}{{\Delta}^2 - x^2}\right)
		& |x| \le \Delta  \vspace{3mm} \\
	 0	& |x| \ge \Delta \, .
	\end{array}
   \right.
\ee
${\s}_{\Delta}$ is symmetric about $x=0$, at which it attains its
maximum value ${\s}_{\Delta}(0) = 1$. Most importantly, ${\s}_{\Delta}$
is an infinitely differentiable function, the support of which is compact.
Putting
${\a}(r, z) := \frac{\pi}{2}{\s}_{\half}(z){\s}_{\half}(r-1)$,
we  obtain \vskip2mm
\noindent $g(r\cos \t, r\sin \t, z) :=
	h_{\frac{\pi}{2}{\s}_{\half}(z){\s}_{\half}(r-1)} (\t)   \vspace{2mm}$
\be
\hspace{24mm}
	= \left(
	\begin{array}{c}
	1 - (1 - \cos\t) \sin^2\!\left[\frac{\pi}{2}\s_{\half}(z)
		\s_{\half}(r-1)\right]
							\vspace{2mm}  \\
	\sin\t \sin\!\left[\frac{\pi}{2}\s_{\half}(z)
		\s_{\half}(r-1)\right]
							\vspace{2mm}  \\
	(1 - \cos\t) \sin\!\left[\frac{\pi}{2}\s_{\half}(z)
		\s_{\half}(r-1)\right]
	\cos\!\left[\frac{\pi}{2}\s_{\half}(z)
		\s_{\half}(r-1)\right]
							\vspace{2mm}  \\
	0
	\end{array}
		\right).					\label{gauge}
\ee
Note that $g$ is infinitely differentiable and is trivial outside a
compact region.  Further, since ${\a}_{\half}(r=1, z=0) = \pi /2$,
its restriction to the circle $\C$ is given by
$g(\cos\t, \sin\t, 0) = \exp(\t{\bf T})$, as desired.

The  purpose of this sub-section was the construction of this
gauge-transformation.  For any connection
of the form written in Eq. (\ref{conn}), the holonomy ``around''
$X_{\c}^t$ is {\em non-covariant}, according to Eq. (\ref{su2}).
For example, choose
\be
	{\bf A}(r\cos \t, r\sin \t, z) :=
	 A{\s}_{\half}(z){\s}_{\half}(r-1) {\bf T} \, d\t, \quad A \!\in \!\R
								\label{A}
\ee
(which also vanishes outside of a compact region).
We then have
\be
  U_A[X_{\c}^t] = \exp[2\pi At{\bf T}],
\ee
and
\be
  U_{A^g}[X_{\c}^t] = \exp[2\pi At{\bf T}] \cdot \exp[2\pi t{\bf T}].
\ee
This completes the first example.

\subsec{Example 2}
The above example involved what may be, from the Lie algebraic point of view,
the most notable element of the extended loop group -- the real power of an
ordinary loop.  The extended loop group also contains elements which
seem radically different than ordinary loops, but which may be quite
useful.  These elements are, in a sense, ``less distributional'' than
the ordinary loops.  The differential relation (\ref{diff}) does not allow
the existence of smooth extended loops, i.e. those for which all entries are
smooth;
they must be genuinely distributional.  However, it is a trivial application
of the results of  \cite{gambini} to show
that given an arbitrary multi-vector density
$Y^{{\mu}_1 \cdots {\mu}_m}$ which is divergence-free in each index
and satisfies the {\em homogeneous} algebraic relation (\ref{hom-alg}),
there exists an element $X \in \sel$ such that $X^{{\mu}_1 \cdots {\mu}_k}=0
\,\,\forall \,\, k<m$ and $X^{{\mu}_1 \cdots {\mu}_m} =
Y^{{\mu}_1 \cdots {\mu}_m}$.
In particular, we may choose $Y^{{\mu}_1 \cdots {\mu}_m}$ to be {\em smooth}.
The set of all extended loops whose first
non-vanishing entry is smooth is a sub-Lie group of $\sel$.  One might
think of these elements as ``smoothened loops''.

As was mentioned in the introduction, the existence of the smoothened loops
is a nice feature of $\sel$.  If $X$ is as described above, then
due to Eq. (\ref{diff}),  $X^{{\mu}_1 \cdots {\mu}_n}$ must  be a
genuine distribution for each $n>m$ .  Therefore, the hope of obtaining a
gauge-invariant smearing of the connection by {\em smooth} functions
is not borne out.
Nonetheless, one might hope \cite{regularization} that some light
may be shed on the problem of regularization of the Wilson loop variables.
Could it be that, by some mathematical miracle, the holonomies of smoothened
loops do not suffer from the problem of non-covariance illustrated above?
By ``smearing out'' the previous example, we will see that the answer to this
question is, unfortunately, negative.

Recall that for Abelian connections, it is only the rank-one entry of $X$
on which the holonomy $U_A[X]$ depends.  Choose an element $X \! \in \! \sel$
for which
\be
  X^{\mu} = \left(\frac{\partial}{\partial \t}\right)^\mu
	{\s}_{\half}(z) \, {\s}_{\half}(r-1).
\ee
$X^\mu$ is a smooth  vector density of compact support,
which may be viewed as a smoothened version of the $X_{\c}^{\mu}$ considered
above.
Let ${\bf A}$ be as in the definition (\ref{A}).  A short calculation
yields the holonomy
\be
  U_A[X] = \exp (X^{\mu}A_{\mu}) = \exp (2\pi A\mbox{e}^6{\bf T}).
\ee

In the spirit of the previous example, we construct a gauge
transformation $\tilde{g}$ which commutes with ${\bf A}$.  The restriction of
$\tilde{g}$ to
the support of ${\bf A}$ must then take values in the Abelian subgroup
generated
by ${\bf T}$.  The idea is simply to replace the smearing function $\s$
in Eq. (\ref{gauge}).
The function
\be
  t_{\D}(x) := \frac{2}{{\D}\mbox{e}^3} \int_{-\infty}^x \! dx' \,{\s}_{\D}(x')
\ee
is a smoothened step function.  It vanishes for all $x \le -\D$ and is
unity for all $x \ge \D$.  Of course, $t_{\D}$ is infinitely
differentiable everywhere.  Define
\be
    s_{\Delta}(x) := \left\{
	\begin{array}{c @ {\quad : \quad} l}
	 t_{\D}(x + 3\D)
		& x \le 0  \vspace{3mm} \\
	 1 - t_{\D}(x - 3\D)
	 	& x \ge 0.
	\end{array}
	\right.
\ee
This function is non-vanishing only for $|x| \le 4\D$ and is constant
on the interval $|x| \le 2\D$, on which it assumes the value one.
Using $s_{\quarter}$ in place of ${\s}_{\half}$ in Eq. (\ref{gauge}),
one obtains the desired gauge transformation;  put \vspace{2mm} \\
$\tilde{g}(r\cos \t, r\sin \t, z) \quad := \quad
  		h_{\frac{\pi}{2}s_{\quarter}(z)s_{\quarter}(r-1)} (\t)$ \\
\be
\hspace{31mm}
	=
	 \left(
	\begin{array}{c}
	1 - (1 - \cos\t)\sin^2\!\left[\frac{\pi}{2}s_{\quarter}(z)
		s_{\quarter}(r-1)\right]
							\vspace{2mm}  \\
	\sin\t \sin\!\left[\frac{\pi}{2}s_{\quarter}(z)
		s_{\quarter}(r-1)\right]
							\vspace{2mm}  \\
	\sin\!\left[\frac{\pi}{2}s_{\quarter}(z)
		s_{\quarter}(r-1)\right]
	(1 - \cos\t) \cos\!\left[\frac{\pi}{2}s_{\quarter}(z)
		s_{\quarter}(r-1)\right]
							\vspace{2mm}  \\
	0
	\end{array}
		\right)	.					\label{gauge2}
\ee
On the support of $X^{\mu}$, $\tilde{g}$ takes a very simple form;  for
$|z| \le \half$  and  $|r-1| \le \half$, \\
$ \tilde{g}(r\cos\t, r\sin\t, z) = (\cos\t, \sin\t, 0, 0) = \exp(\t{\bf T})$.
One may then obtain
\be
 \int X^{a} \, \tilde{g}^{-1}({\bf d}\tilde{g})_a =
				\frac{\pi \mbox{e}^3}{8} {\bf T}.
\ee
Finally,
\be
  U_{A^{\tilde{g}}}[X] =
	U_A[X] \cdot \exp \left( \frac{\pi \mbox{e}^3}{8} {\bf T} \right).
\ee
The extended ``holonomies'' of the smoothened loops are  not covariant
with respect to small gauge transformations.

\sec{Generalizations and conclusions}
The extended loop group is a  well-defined mathematical object.
It is an infinite-dimensional group  which encompasses
the group of based loops on an arbitrary connected manifold, $\M$.
 (Note that we have used the term ``Lie group''
 fairly loosely.  For the sake of rigor, it should be shown that $\sel$
 admits a manifold structure with respect to which the group operations
 are continuous.)
For applications to physics, however, one would also like
to extend the concept of holonomy.  In fact, the construction of the
extended loop group was based on the functional form of the holonomy of
ordinary loops.  There is then the obvious candidate for a generalized
holonomy.  We have found, however, that for the case ${\M} \approx {\R}^3$
this generalized holonomy is
not covariant with respect to small gauge transformations.
Its trace  does  not provide
gauge-invariant functionals on the space of connections for
an $SU(2)$ gauge theory on Minkowski space, for example.

In fact, the result is of a
very general validity.  Since any simple Lie group contains an $SU(2)$
subgroup, it extends to the non-Abelian case with such
gauge groups -- those which are typically relevant in physics.
The result also applies to the case of gravity in terms of the
Ashtekar variables.
Further, although we restricted our attention to ${\M} \approx {\R}^3$, all
of the mappings used in the first example are of compact support.
We may then extend the result to an arbitrary manifold.
(Note, however, that since the topology of ${\R}^3$ was used in a critical
way in defining the smoothened loop group, the second example does not
extend to manifolds of arbitrary topology; i.e. it is not clear that one can
even define the smoothened loop group for the arbitrary case.)

{}From our point of view, the potential
power of the extended loop group involves the use of the traced holonomies as
a large class of observables for gauge theories.  Our results then suggest
a re-evaluation of the extended loop group as an arena for quantum gravity
and Yang-Mills theory.

There are three alternatives worth consideration.  First, the following
question arises:  What characterizes those generalized loops for which the
holonomies are covariant?  Perhaps consideration of this question would
shed some light on the appropriate extension of the loop representation.
  Note however, that by Example 1, one will not have the continuous
structure of a Lie group at one's disposal.  Thus, techniques involving
functional differentiation are not likely to be straight-forward in
such a formulation.   A second alternative, suggested by Gambini
and Pullin \cite{gam-pul}, is to design a different extension of the
holonomy which
is covariant. Alternatively, since  observables of the theory are
our primary concern, it may be most productive to
focus on an extension of the concept of the Wilson loops.  It may be the
case, for example, that such an extension exists which does not manifest itself
as the trace of a holonomy.  Lastly, while it seems that some generalization
of the ordinary loop representation is needed, it may turn out that the
appropriate generalization is altogether different than that suggested
by the existence of the extended loop group.


\vskip4mm
\noindent {\it Acknowledgements:}  I am very grateful to Abhay Ashtekar,
Rodolfo Gambini, and Jorge Pullin for many comments and suggestions.
This work was supported in part by the NSF Grant PHY93-96246 and the
Eberly research fund of The Pennsylvania State University.

\pagebreak

\end{document}